\documentclass[aps,prl,reprint,showpacs,superscriptaddress,nofootinbib,twocolumn]{revtex4-2}
\usepackage{graphicx}
\usepackage{subfigure}
\usepackage{booktabs}
\usepackage{graphicx}
\usepackage{upgreek}
\usepackage{subfigure}
\usepackage{soul}
\usepackage{color}
\usepackage{amsmath}
\usepackage[percent]{overpic}
\begin{document}
\newcommand\mean[1]{\ensuremath{\langle #1\rangle}}
\newcommand\ket[1]{\ensuremath{|#1\rangle}}
\newcommand\bra[1]{\ensuremath{\langle#1|}}

\newcommand{\y}[1]{{\color{blue} #1}}
\newcommand{\yc}[1]{{\color{red} #1}}
\newcommand{\yr}[2]{{\color{blue}{\st{#1}} #2}}
\newcommand{\yd}[1]{{\color{blue} \st{#1}}}

\newcommand{\jc}[1]{{\color{green} #1}}

\title{\textbf{Experimental Multi-Dimensional Side-Channel-Secure  Quantum Key Distribution}}

\author{Hao Dong}
\affiliation{Hefei National Research Center for Physical Sciences at the Microscale and School of Physical Sciences,
University of Science and Technology of China, Hefei, Anhui 230026, China}
\affiliation{Jinan Institute of Quantum Technology and Hefei National Laboratory Jinan Branch, Jinan, Shandong, China}

\author{Cong Jiang}
\affiliation{Jinan Institute of Quantum Technology and Hefei National Laboratory Jinan Branch, Jinan, Shandong, China}
\affiliation{Hefei National Laboratory, University of Science and Technology of China, Hefei, Anhui 230088, China}

\author{Di Ma}
\affiliation{Jinan Institute of Quantum Technology and Hefei National Laboratory Jinan Branch, Jinan, Shandong, China}

\author{Chi Zhang}
\affiliation{Jinan Institute of Quantum Technology and Hefei National Laboratory Jinan Branch, Jinan, Shandong, China}

\author{Jia Huang}
\affiliation{National Key Laboratory of Materials for Integrated Circuits, Shanghai Institute of Microsystem and Information Technology, Chinese Academy of Sciences (SIMIT, CAS)}

\author{Hao Li}
\affiliation{National Key Laboratory of Materials for Integrated Circuits, Shanghai Institute of Microsystem and Information Technology, Chinese Academy of Sciences (SIMIT, CAS)}

\author{Li-Xing You}
\affiliation{National Key Laboratory of Materials for Integrated Circuits, Shanghai Institute of Microsystem and Information Technology, Chinese Academy of Sciences (SIMIT, CAS)}

\author{Yang Liu}
\affiliation{Hefei National Research Center for Physical Sciences at the Microscale and School of Physical Sciences,
University of Science and Technology of China, Hefei, Anhui 230026, China}
\affiliation{Jinan Institute of Quantum Technology and Hefei National Laboratory Jinan Branch, Jinan, Shandong, China}
\affiliation{Hefei National Laboratory, University of Science and Technology of China, Hefei, Anhui 230088, China}

\author{Xiang-Bin Wang}
\affiliation{Jinan Institute of Quantum Technology and Hefei National Laboratory Jinan Branch, Jinan, Shandong, China}
\affiliation{Hefei National Laboratory, University of Science and Technology of China, Hefei, Anhui 230088, China}
\affiliation{State Key Laboratory of Low Dimensional Quantum Physics, Department of Physics, Tsinghua University, Beijing 100084, China}

\author{Qiang Zhang}
\affiliation{Hefei National Research Center for Physical Sciences at the Microscale and School of Physical Sciences,
University of Science and Technology of China, Hefei, Anhui 230026, China}
\affiliation{Jinan Institute of Quantum Technology and Hefei National Laboratory Jinan Branch, Jinan, Shandong, China}
\affiliation{Hefei National Laboratory, University of Science and Technology of China, Hefei, Anhui 230088, China}

\author{Jian-Wei Pan}
\affiliation{Hefei National Research Center for Physical Sciences at the Microscale and School of Physical Sciences,
University of Science and Technology of China, Hefei, Anhui 230026, China}
\affiliation{Hefei National Laboratory, University of Science and Technology of China, Hefei, Anhui 230088, China}

\date{\today}

\begin{abstract}
Quantum key distribution (QKD) theoretically provides unconditional security between remote parties. However, guaranteeing practical security through device characterisation alone is challenging in real-world implementations due to the multi-dimensional spaces in which the devices may be operated. The side-channel-secure (SCS)-QKD protocol, which only requires bounding the upper limits of the intensities for the two states, theoretically provides a rigorous solution to the challenge and achieves measurement-device-independent security in detection and security for whatever multi-dimensional side channel attack in the source.  Here, we demonstrate a practical implementation of SCS-QKD, achieving a secure key rate of $6.60$ kbps through a 50.5 km fibre and a maximum distribution distance of 101.1 km while accounting for finite-size effects. Our experiment also represents an approximate forty-times improvement over the previous experiment.
\end{abstract}

\maketitle

\section{Introduction} 
In theory, the quantum key distribution (QKD) generates secure keys that are information-theoretically secure~\cite{RevModPhys.92.025002}, even if hackers have the most powerful attacks that physical laws permit. However, in practical implementations, components may deviate from the protocol requirements, introducing vulnerabilities that eavesdroppers (Eve) could exploit to compromise system security~\cite{PhysRevA.78.042333,Lydersen2010,PhysRevLett.112.070503,Yoshino2018,PhysRevApplied.10.064062}. As a result, Eve could steal the secure keys without being discovered. Therefore, when designing QKD protocols, the imperfections of the practical components should be considered to guarantee practical security.

Measurement-device-independent (MDI) QKD~\cite{PhysRevLett.108.130503,RevModPhys.92.025002} protocols eliminate all detection-side loopholes from theory, significantly improving practical security. Recent developments in twin field (TF) QKD~\cite{Lucamarini2018,PhysRevLett.123.100505,PhysRevLett.124.070501,Chen2021,PhysRevLett.128.180502,PhysRevLett.130.210801} and mode-pairing QKD~\cite{Zeng2022,PhysRevLett.130.030801,Zhu:24} are MDI-type QKD protocols and thus inherit high practical security. The decoy-state method~\cite{PhysRevLett.91.057901,PhysRevLett.94.230503,PhysRevLett.94.230504} was developed to counteract photon-number-splitting (PNS) attack~\cite{PhysRevLett.85.1330,PhysRevA.61.052304} targeting imperfect single-photon sources. This method enables practical secure QKD implementation using weak coherent pulses generated by attenuated lasers. 

However, loopholes may still exist in the source. The side channel effects in multi-dimensional space, including the frequency spectrum, the emission time, the shape of the wave, etc., can leak extra information to the eavesdropper, compromising the security of practical QKD systems~\cite{PhysRevA.98.012330,RevModPhys.92.025002}. Impressively, a type of such effect in multidimensional space named time-varying encoding was experimentally demonstrated and has shown its potential to undermine the security of source sides in QKD~\cite{gnanapandithan2025hidden}. There are several ways to solve the threat in practice. For example, we can set standards for QKD devices against all known attacks and make well calibration before deployment, but it might be impossible to defend all multi-dimensional side channel attacks. 

The side channel secure (SCS) QKD protocol~\cite{PhysRevApplied.12.054034} is theoretically proven secure over the whole-space states, including side channels in multi-dimensional space and operational space. The original SCS-QKD protocol~\cite{PhysRevApplied.12.054034} assumes perfect vacuum states and requires that the coherent-state intensities be upper-bounded to ensure security. Recent theoretical advances have addressed the imperfect vacuum~\cite{jiang2023side} and established a method to calculate the secure key rate considering finite-key length~\cite{PhysRevResearch.6.013266}. The only assumptions are that Eve cannot access inside Alice's or Bob's laboratories, and the source states of each time window prepared inside the laboratory are independent of state choice information of other time windows. The real SCS-QKD protocol with imperfect sources can be mapped to a perfect protocol with two perfect sources, the secure key rate can be calculated as if we were carrying out the perfect protocol. Importantly, through worst-case estimation for the real states in Fock space, the SCS protocol does not need the side-channel information about states in the multi-dimensional side channel space and thus is immune to any attacks in the multi-dimensional space.

In this work, we experimentally demonstrated the SCS-QKD protocol~\cite{PhysRevApplied.12.054034,PhysRevResearch.6.013266} across 0 km to 101.1 km fibre distances. Independent laser sources are stabilized locally using saturated absorption spectroscopy of acetylene as an absolute frequency reference, and real-time active phase compensation is used to improve the secure key rate. We achieve a finite-size secure key rate (SKR) of 64.7 kbit/s at 0 km and 6.6 kbit/s at 50 km, which are approximately 40 times higher than the rates reported in previous work~\cite{PhysRevLett.128.190503}. This work represents a practical implementation by eliminating the ideal vacuum and asymptotic assumptions required in the previous experiment.

\begin{figure*}
\includegraphics[width=0.8\linewidth]{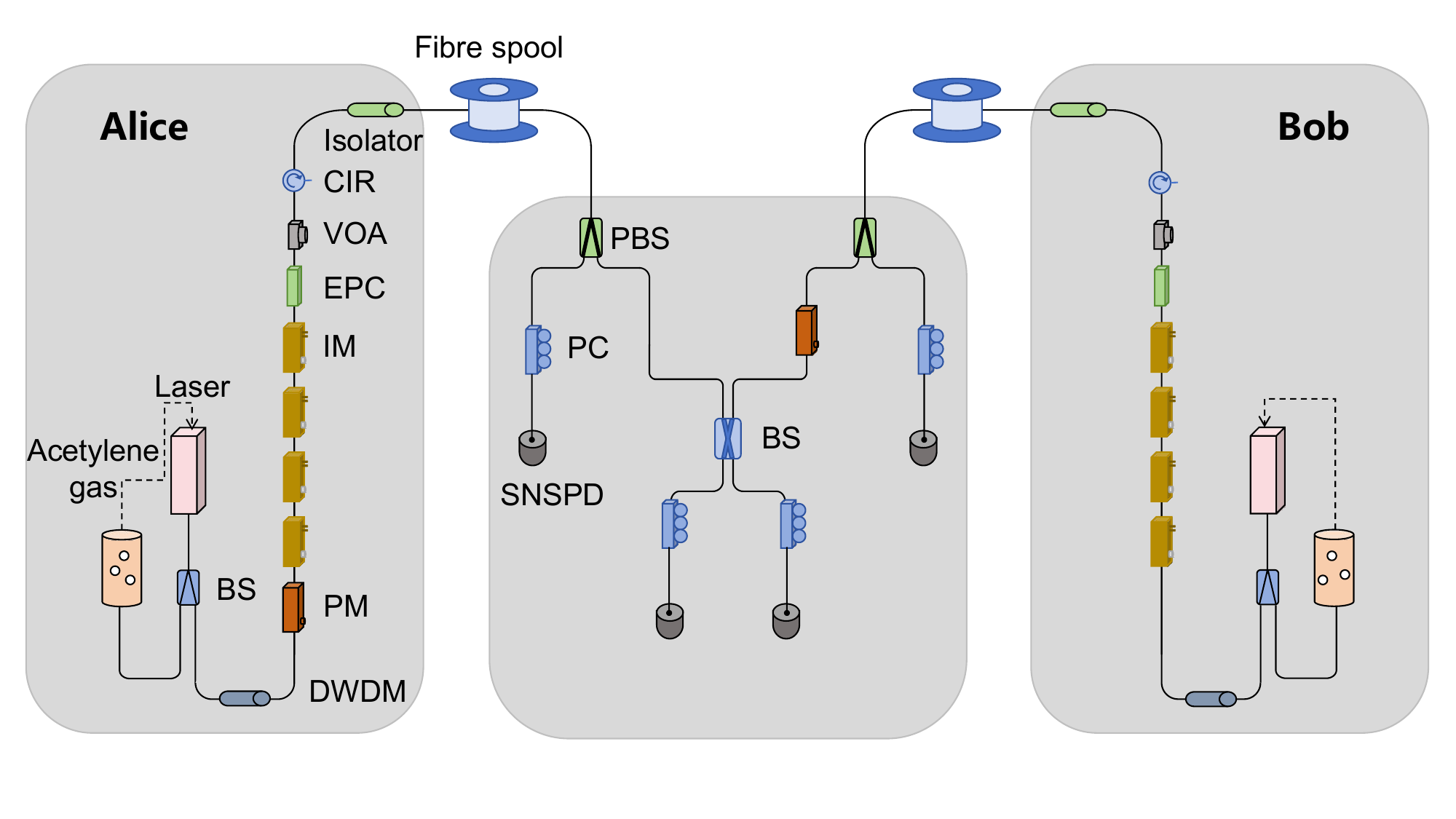}
\caption{\label{fig1}Experimental setup. In Alice’s and Bob’s lab, two acetylene-stabilized lasers are employed as the light sources. First, these lasers are filtered by a dense wavelength division multiplexer (DWDM). Then they are modulated with four intensity modulators (IMs) (and a phase modulator (PM) in Alice's laboratory) to generate a waveform pattern that time multiplexes the weak signal pulses with strong phase reference pulses and attenuates them to bring the signal pulses to the single-photon level with an attenuator (ATT). A circulator (CIR) and an optical isolator are placed immediately following the attenuator to prevent injection attacks. The prepared light pulses are finally sent to Charlie through the ultralow-loss fibre spools for detection. Charlie interferes and measures the light from Alice and Bob with superconducting nanowire single-photon detectors (SNSPDs). EPC, electric polarization controller; PC, polarization controller.}
\end{figure*}

\section{Protocol}

Here, the practical two-state SCS-QKD protocol~\cite{PhysRevApplied.12.054034,PhysRevResearch.6.013266} with phase reference pulses and an active phase compensation method is used.

The protocol involves the following main steps. Alice and Bob first randomly prepare vacuum or coherent states with specific probabilities, which are called signal pulses. The prepared vacuum or coherent states can be imperfect in the whole space and different in different time windows. Alice (Bob) also prepares a strong reference pulse, time-multiplexed with the signal pulses. The phases of the reference pulses are modulated periodically, and this will be presented in detail later. No matter what she (he) decides, she (he) always sends the reference pulse to Charlie. 

After receiving the signal pulses, Charlie first performs phase compensation and then an interferometry measurement at his measurement station. The measurement results will be announced to Alice and Bob. We define the effective window as the event where the right-side detector clicks and the left-side detector does not click. 

After repeating the above process $N$ times, Alice and Bob use data from effective windows for the raw bits. In the data postprocessing, they first perform error correction, and then they calculate the secure final key rate under collective attack by

\begin{equation}
\label{keyrate_col}
\begin{aligned}
    R_{col}  =  &\frac{1}{N}  \big\{ n_{\mathcal{Z}}  [ 1-H(\bar{e}  ^{ph})  ]  - leak_{EC}  -  \log_{2}{\frac{2}{\varepsilon  _{cor}}}\\
     &- 2\log_{2}{\frac{1}{\varepsilon  _{PA}}}  - (d+3)\sqrt{n_{\mathcal{Z}}\log_{2}{\frac{2}{\bar{\varepsilon} } }}  \big\}.
\end{aligned}
\end{equation}
   
Here, $leak_{EC}$ is the amount of information leakage during the error-correction process, and generally $leak_{EC} = fM_{s}H(E_{\mathcal{Z}})$, where $f$ is the error-correction inefficiency, $M_{s}$ is the number of raw keys, and $E_{\mathcal{Z}}$ is the bit-flip error rate of the raw key strings. $n_{\mathcal{Z}}$ is the number of effective windows in which only one of Alice and Bob decides to send out a coherent state pulse. $H(x)=-x\log_{2}{x}-(1-x) \log_{2}{(1-x)}$ is the Shannon entropy, $\varepsilon _{cor}$ is the failure probability of the error correction, $\varepsilon _{PA}$ is the failure probability of the privacy amplification, $\bar{\varepsilon }$ is the coefficient for measuring the accuracy of estimating the smooth min-entropy, $d$ is the dimension of the local states shared by Alice and Bob, where $d= 8$ in the SCS protocol. With Eq.~(\ref{keyrate_col}), the protocol is $\varepsilon_{col}=\varepsilon_{cor}+\bar{\varepsilon}+\varepsilon_{PA}+3\epsilon$ secure against collective attacks, where $\epsilon$ is the failure probability in using statistical fluctuation.

Applying the postselection technique, by shortening $2(d^{2}-1)\log_{2}{(N+1)}$ bits of the final key distilled under collective attack, we can get secure final keys against the coherent attack with the security coefficient $\varepsilon _{coh}=\varepsilon _{col}(N+1)^{d^2-1}$. Finally, we can obtain the key rate formula under coherent attack by
\begin{eqnarray}\label{keyrate_coh}
R_{coh}=R_{col}-\frac{2(d^{2}-1)\log_{2}{(N+1)})}{N}.
\end{eqnarray}

In this work, we set $\varepsilon _{coh}=10^{-10}$ and $\varepsilon_{cor}=\bar{\varepsilon}=\varepsilon_{PA}=\epsilon$.

\section{Experiment}

\begin{figure}
\includegraphics[width=\linewidth]{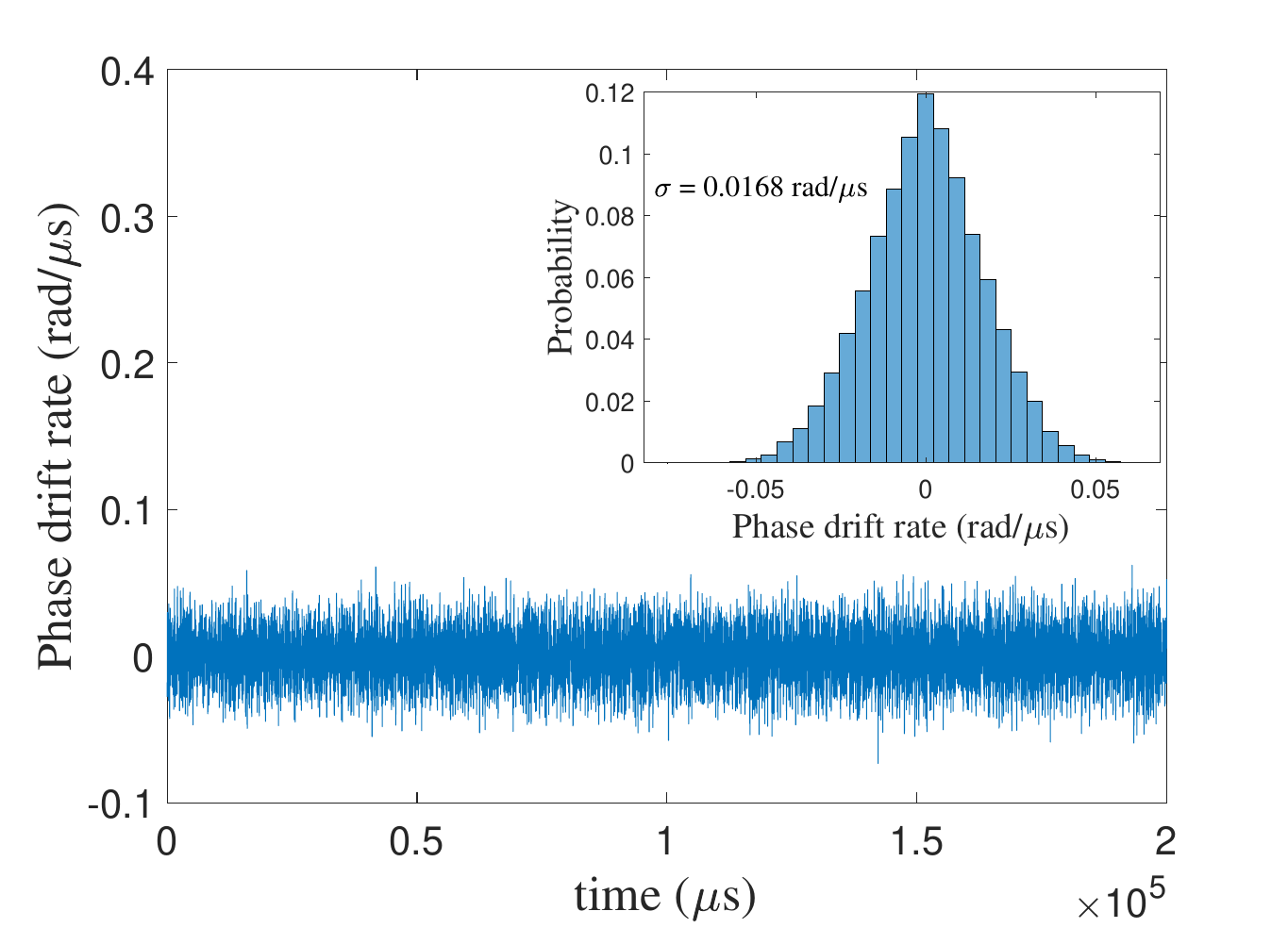}
\caption{\label{fig2}Relative phase drift rate between the light sources. The interference signals from the two light sources were captured using an oscilloscope operating at a sampling rate of 1 MSa/s, with a total acquisition duration of 200 ms. A moving average filter with a 25 $\upmu$s window was applied to suppress high-frequency electrical noise. The standard deviation of the relative phase drift between the two light sources is 0.0168 rad/$\upmu$s.}
\end{figure}

The experimental setup is shown in Fig.~\ref{fig1}, consisting of the legitimate users (Alice and Bob) and the measurement node (Charlie). In the laboratories of Alice and Bob, independent fibre lasers serve as the light source without any external wavelength reference. The lasers are locked to specific transitions of acetylene molecules, with a central wavelength of 1542.3837 nm. We measured the phase fluctuation rate to be approximately 0.0168 rad/$\upmu$s between the two independent lasers (see Fig.~\ref{fig2}). We attribute this phase drift to the different environmental conditions of the acetylene cells. The acetylene-stabilized lasers ensure highly stable frequency precision without external input reference, which is typically required in an optical phase-locked loop (OPLL) system. Thereby, the method avoids potential loopholes associated with external light injection~\cite{Peng2025}.

The light is then modulated using a series of three intensity modulators.
The SCS-QKD protocol encodes information solely through the ``sending'' and ``not sending'' signals. Three cascaded intensity modulators are used to achieve the ultra-low vacuum intensity required. The encoded signal is then attenuated to single-photon level at the output. Strong reference pulses are time-division multiplexed with the signal pulses to compensate for the relative phase drift~\cite{PhysRevLett.123.100505}. (See Supplementary Materials for details of the signal generation.) To prevent injection attacks~\cite{PhysRevApplied.20.044005,PhysRevX.5.031030}, e.g., the laser-seeding attack (LSA), a circulator and an optical isolator are placed immediately following the attenuator. The total isolation of the isolator, circulator, and the attenuator exceeds 169 dB.

The signals are then transmitted to Charlie’s measurement station via fibre spools with a total length from 0 km to 101.1 km. Upon arrival, the light is first polarized using a polarization beam splitter (PBS). One output of the PBS is directly routed to superconducting nanowire single photon detector (SNSPD), serving as the monitoring port; the other output serves as the signal port, is then directed to a beam splitter (BS) for interference. From the monitoring port, the detection count is stabilized to approximately 10\% of the maximum intensities, by adjusting the electrically controlled polarization controller (EPC) located at Alice's (Bob's) side, to compensate for the polarization drift induced by the fibre channel. Consequently, the intensities of the signal port remain stable for subsequent interference. 

\begin{figure}[t]
\centering
\subfigure {\
\begin{minipage}[t]{\linewidth}
        \centering
            \begin{overpic}[scale=0.5]{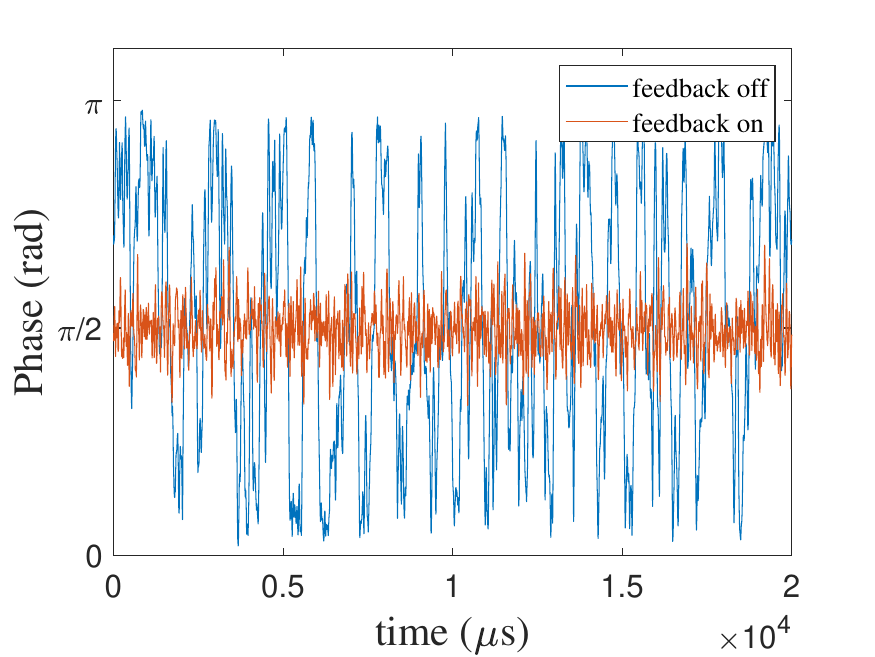}
            \put(1,70){(a)}
        \end{overpic}
    \end{minipage}
    \label{subfig3a}
}
\subfigure {\
\begin{minipage}[t]{\linewidth}
        \centering
           \begin{overpic}[scale=0.5]{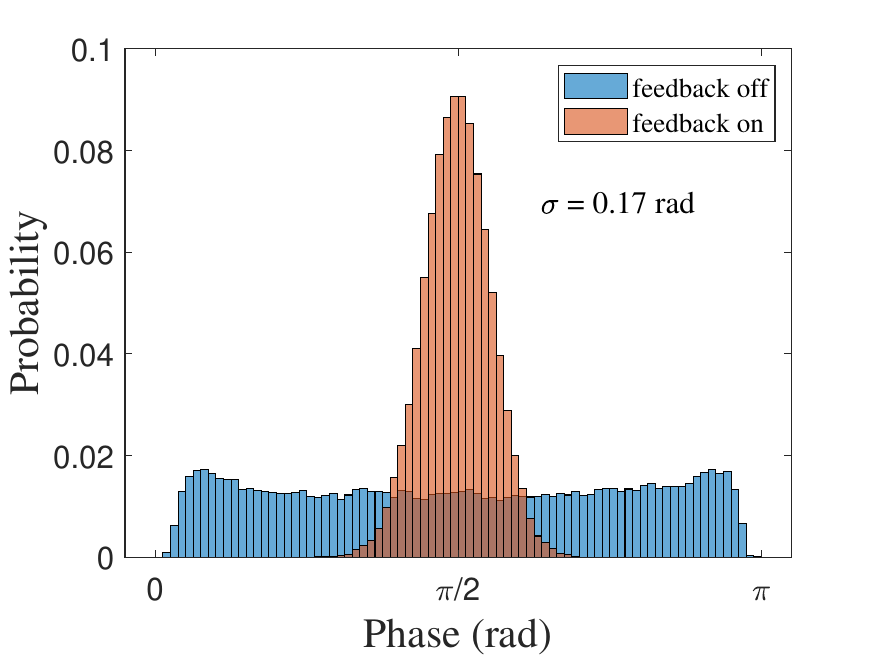}
           \put(1,70){(b)}
        \end{overpic}
    \end{minipage}
    \label{subfig3b}
}
\label{fig3}
\caption{ (a) Phase drift compensation with the feedback on and off. The optical output power as a function of time is measured at one output of Charlie's beam splitter is shown. (b) Phase distribution with the feedback on and off. The standard deviation of the phase fluctuation is 0.88 rad when the feedback is off, and 0.17 rad when the feedback is on.}
\end{figure}

The fast phase fluctuation induced by the channel fibre is compensated in real-time with a PM positioned at the signal port of Bob's PBS. A custom-designed field-programmable gate array (FPGA) board was developed to execute the feedback. The proportional-integral-derivative (PID) algorithm was configured to stabilize the phase difference of the channel to zero. In all experimental trials, the intensity of the reference light is set to approximately 2.5 MHz detections for the feedback. (See Supplementary Materials for details of the real-time feedback design.) The performance of the real-time compensation is evaluated by comparing the interference results with the feedback on and off. As shown in Fig.~\ref{subfig3a} and Fig.~\ref{subfig3b}, the standard deviation of the phase fluctuation is suppressed to 0.17 rad with the real-time feedback. 

\section{result}

Our system operates at a repetition rate of 250 MHz with a 700 ps signal pulse width. The quantum signals are emitted in the last 40 ns of the 100 ns cycle, resulting in an effective signal frequency of 100 MHz. The extinction ratio between the signal and vacuum states is higher than 70 dB for all experimental trials. The standard deviation of the signal intensity fluctuation is approximately 0.54\% of the average signal intensity. The interference results are measured with two SNSPDs and recorded by a time tagger. The detection efficiencies of the SNSPDs are 68.2\% and 70.5\%, with dark count rates of 8.4 Hz and 7.5 Hz, respectively. (See Supplementary Materials for details of the parameters).

\begin{figure}[h]
\includegraphics[width=\linewidth]{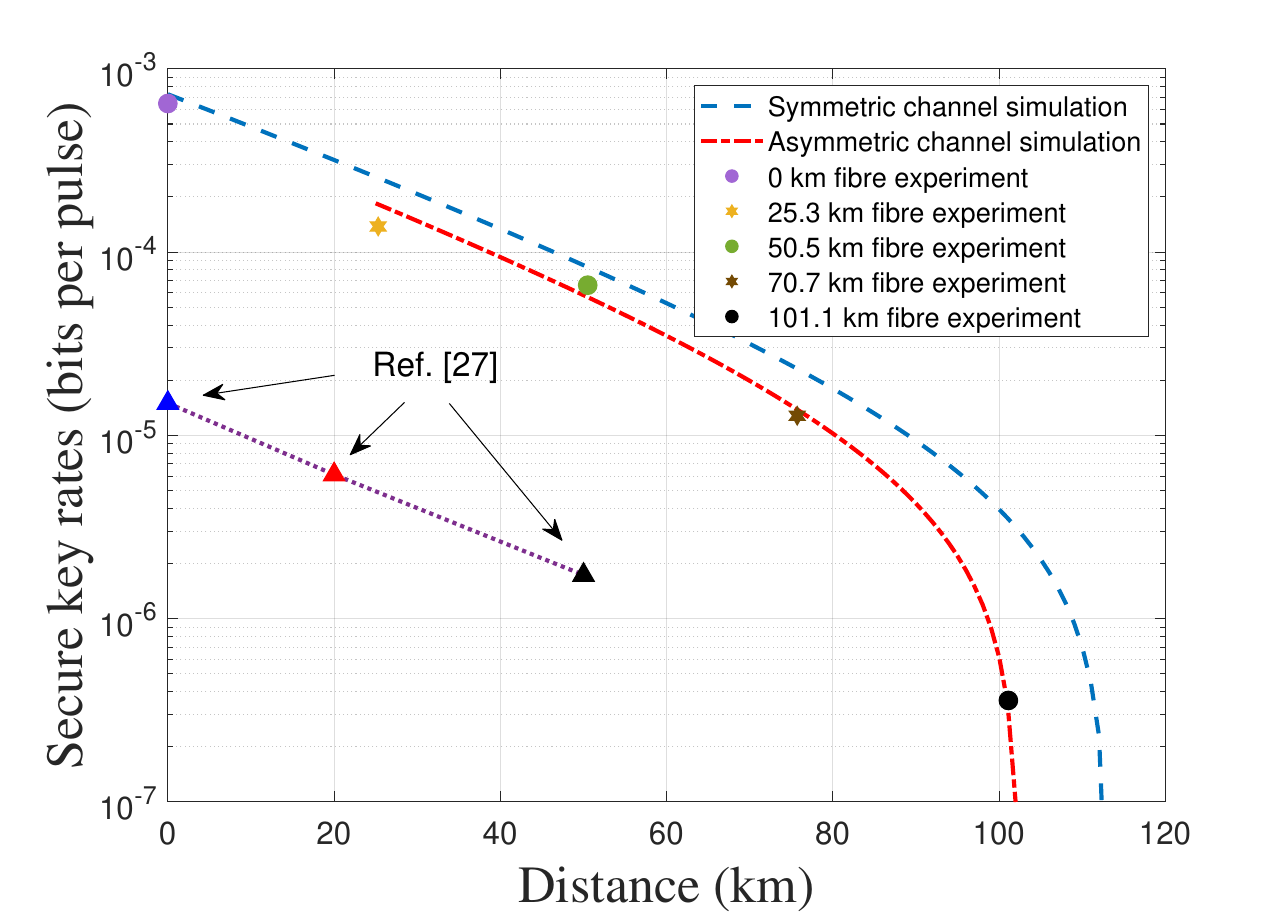}
\caption{\label{fig4}Simulated and experimental secure key rates. Experimental results are shown for (1) symmetric fibre lengths: 0 km (purple dot), 50.5 km (green dot), and 101.1 km (black dot); (2) asymmetric fibres: 25.3 km (yellow star) and 70.7 km (brown star). Theoretical simulations are illustrated as dashed curves: blue for optimized symmetric channels and red for optimized asymmetric channels. Previous experimental SCS-QKD results from Ref.~\cite{PhysRevLett.128.190503} are shown as triangles for comparison.}
\end{figure}

We experimentally demonstrated SCS-QKD over fibre lengths from 0 km to 101.1 km between Alice and Bob. The average attenuation of the fibres is 0.168 dB/km. The finite size effect is taken into consideration for all experimental tests to ensure the composable security under any coherent attack. In the calculation, the upper bound of the signal intensity is set to 102.15\% of the average signal intensity, considering 4 times the standard deviation of the signal fluctuation. The error correction inefficiency is set to $f=1.16$, the statistics block size is set to the complete length of the test, the security coefficient under coherent attack $\varepsilon _{coh} = 10^{-10}$.

For the 0 km and 50.5 km cases, we set the fibre length of the Alice-Charlie and Bob-Charlie paths to be the same. However, the additional PM in Bob's path introduced a 3.05 dB higher insertion loss for Bob's path. To compensate for this imbalance, we optimized the transmitter's optical intensity (See Supplementary Materials for details of the parameters) to maximize the key rate. The total number of pulses sent by Alice and Bob is set to $3.63\times 10^{11}$, corresponding to 1.01 h experimental time. The achieved secure key rate is 64.7 kbps for 0 km and 6.60 kbps for 50.5 km, as summarized in Fig.~\ref{fig4}. The quantum bit error rates (QBER) are 1.95\% and 2.47\%, respectively. Note that these secure key rates represent a $\sim40\times$ improvement over the previous SCS-QKD experiment~\cite{PhysRevLett.128.190503}. 

For the 25.3 km and 75.7 km cases, we set the fibre length asymmetry to be approximately 25 km, and placed a 1.2 dB attenuator on Bob's side to balance the total loss, including Charlie's optics for Alice's and Bob's paths. The total number of pulses sent is $3.63\times 10^{11}$, the same as in the 0 km and 50.5 km case. The achieved secure key rate is 13.8 kbps for 25.3 km and 1.28 kbps for 75.7 km, with QBERs of 2.99\% and 2.48\%. 

For the 101.1 km fibre length, the fibre length is set to be symmetric. Additionally, we set a 600 ps-width detection window to filter noise during data processing. This retains approximately 79\% of the detections as valid events, decreasing the effective detection efficiency to about 55\%. The number of quantum pulses sent is increased to $1.126\times 10^{12}$, enabling us to collect a statistically reliable measurement dataset for evaluating finite-size effects. The achieved secure key rate is 35.8 bps.

In conclusion, we have experimentally demonstrated the SCS-QKD protocol. Compared with the previous experiment, our work addresses security, considering imperfect vacuum states and coherent attacks with finite pulse numbers. The results of our experiment are secure under whatever side channels in multi-dimensional space, such as the frequency spectrum, the emission time and so on. The experiment achieves superior key rates and extends the maximum secure transmission distance to over 100 km. The achieved SKR below 100 km is comparable to that of previous MDI-QKD experiments~\cite{RevModPhys.92.025002}; the SKR is about 15–35\% of the theoretically optimized MDI-QKD when intensity fluctuations are neglected and is comparable to MDI-QKD when considering the same level of intensity fluctuations.

This work was supported by the Key R\&D Plan of Shandong Province (Grant No. 2021ZDPT01), Innovation Program for Quantum Science and Technology (2021ZD0300700), the National Natural Science Foundation of China (Grant Nos. T2125010, 12174215 and 12374470), Shandong Provincial Natural Science Foundation Grant No. ZR2021LLZ007, the Chinese Academy of Sciences, C.J., X.-B.W. and Q.Z.  acknowledge support from the Taishan Scholar Program of Shandong Province, Q.Z. was supported by the New Corner Stone Science Foundation through the Xplorer Prize.

\bibliography{ref}
\clearpage

\section{Supplementary materials}

\subsection{The protocol and calculation method}
Our experiment has adopted one form of SCS QKD protocol~\cite{PhysRevApplied.12.054034,jiang2023side,PhysRevResearch.6.013266} using a supposed coherent state and a supposed vacuum. Below we show the protocol and calculation formula of secure key rate according to Ref.~\cite{PhysRevResearch.6.013266}.

\textbf{Main idea.} In a perfect two-state protocol, two coherent states (one of them can be vacuum) are prepared exactly and we only need to write the states in the operational space, i.e. the photon number space. There are explicit formulas to upper bound the phase error and calculate the secure key rate. The real SCS-QKD protocol with imperfect sources can be mapped from a perfect protocol with two perfect sources~\cite{PhysRevApplied.12.054034,jiang2023side,PhysRevResearch.6.013266} by map $\mathcal{M}$ and the secure key rate can be calculated as if we were carrying out the perfect protocol. Since the mapping can in principle be done by channel controlled by Eve, and hence the real protocol must be secure if the perfect protocol is secure. Otherwise, if the real protocol were insecure under Eve’s attack $\mathcal{A}$, the perfect protocol must be insecure under Eve’s attack $\mathcal{A}\mathcal{M}$. Importantly, by the worst-case estimation for the real states in Fock space, the SCS-QKD protocol does not need the side-channel information of states in the multi-dimensional side channel space.

The protocol involves the following three steps:

\textbf{Step1.} In the \textit{i}th time window, Alice (Bob) randomly commits to classical bit value 0 or 1 with probabilities $p_{0}$ and $p_{x}=1-p_{0}$ ($p_{x}$ and $p_{0}$). Alice (Bob) wants to prepare a perfect coherent state for bit value 1 (0) and a perfect vacuum for bit value 0 (1), but the real source can only produce an imperfect coherent states and imperfect vacuum states. Specifically, when Alice (Bob) wants to prepare a perfect coherent state $x_A$ ($x_B$), she (he) actually prepares a state in 
\begin{equation}\label{real_statep2}
\begin{split}
&\ket{x _{Ai}}=\sqrt {a_{0i}}  \ket{0}+\sqrt{1- a_{0i} } \ket{\tilde \psi_{Ai}},\\
&\ket{x _{Bi}}=\sqrt {b_{0i}} \ket{0}+\sqrt {1- b_{0i}} \ket{\tilde \psi_{Bi}}.
\end{split}
\end{equation}
Here $\ket{0}$ is the vacuum state and $\ket{\tilde \psi_{Ai}}$ ($\ket{\tilde \psi_{Bi}}$) is a whole-space state containing at least 1 photons. If Alice (Bob) wants to prepare a vacuum state $o_A$ ($o_B$), she (he) actually prepares a state in
\begin{equation}\label{real_state1}
\begin{split}
&\ket{o_{Ai}}=\sqrt{a_{v0i}}\ket{0}+\sqrt{1-a_{v0i}}\ket{\tilde{ \phi}_{Ai}},\\
(&\ket{o_{B{i}}}=\sqrt{b_{v0i}}\ket{0}+\sqrt{1-b_{v0i}}\ket{\tilde{\phi}_{Bi}}).
\end{split}
\end{equation}
Here state $\ket{\tilde {\phi}_{Ai}}$ ($\ket{\tilde {\phi}_{Bi}}$) is a whole-space states containing at least 1 photons.
For all time windows in the protocol, the following condition is satisfied
\begin{align}
    &e^{-\mu_0^U}\le a_{v0i}, \quad  e^{-\mu_0^U}\le b_{v0i},\\
    &e^{-\mu_A^U}\le a_{0i}, \quad  e^{-\mu_B^U}\le b_{0i}.
\end{align}

These pulses are called signal pulses. Alice (Bob) also prepares a strong reference pulse time-multiplexed with the signal pulses. The phases of the reference pulses are modulated periodically and this will be presented in detail later. They define $\mathcal{O}$ ($\mathcal{B}$) windows as time windows when Alice and Bob both decide on sending vacuum (coherent states), and define $\mathcal{Z}$ windows as time windows when only one of Alice or Bob decides on the coherent state, the bits of $\mathcal{Z}$ windows are untagged bits.

Any real quantum state lives in the whole space rather than in the operational space only. Here we consider effects of imperfections of the whole-space states, including the side channel effects which are caused by state imperfections beyond the operational space and the state error in the operational space. For example, consider the BB84 protocol using photon polarization. The real state lived in the whole space instead of polarization space only. There errors in the whole space state includes the imperfections beyond the polarization space, such as the imperfections in frequency spectrum, in emission time and so on. Of course there are also  errors in the operational space itself, i.e., the polarization states produced by the real set-up are a bit different from the exact BB84 states.

\textbf{Step2.} After received signal pulses, Charlie first performs phase compensation and then interferometry measurement at his measurement station. The measurement results would be announced to Alice and Bob. In the SCS protocol, most of the bit-flips come from $B$ windows, where both Alice and Bob have chosen to send out the coherent states. The key rate can be greatly improved if Alice and Bob only consider the clicking events caused by the right detectors. Thus here, if the right-side detector clicks and the left-side detector does not click, it is regarded as an effective window.

\textbf{Step3.} Alice and Bob use data from effective windows for the raw bits. In the data postprocessing, they first take error correction. After this, Alice and Bob shall know the values of $n_{\zeta }$, ($\zeta$ = $\mathcal{O}$, $\mathcal{B}$, $\mathcal{Z}$), where $n_{\zeta }$ is the number of effective $\zeta$ windows.

As shown in Ref.~\cite{PhysRevResearch.6.013266}, the real protocol can be mapped from a perfect two-state protocol where the two states are perfect vacuum and perfect coherent states respectively. Thus $\bar{e}^{ph}$ can be calculated by the following formulas With the values of $n_{\zeta }$.

\begin{equation}
\begin{split}
  \left \langle  N^{ph}\right \rangle \le \left \langle \bar{N}^{ph}\right \rangle &=\frac{p_0p_x}{2}\left[\frac{c_0^2}{p_0^2}\left \langle  n_{\mathcal{O}}\right \rangle^U+\frac{c_1^2} {p_x^2}\left \langle n_{\mathcal{B}}\right \rangle^U +\bar{c}_2^2N\right.\\ 
  &\left.+ \frac{2c_0c_1}{p_0p_x}\sqrt{\mean{n_{\mathcal{O}}}^U\mean{n_{\mathcal{B}}}^U}+ \frac{2c_0\bar{c}_2}{p_0}\sqrt{N\mean{n_{\mathcal{O}}}^U}\right.\\
  &\left.+\frac{2c_1\bar{c}_2}{p_x}\sqrt{N\mean{n_{\mathcal{B}}}^U} \right],  
\end{split}
\end{equation}

where $\mean{N^{ph}}$ is the expected value of the number of phase errors and $\mean{\bar{N}^{ph}}$ is its upper bound; $c_0=c_1=1$ and
\begin{equation}
\begin{aligned}
   \bar{c}_2&=\sqrt{ [c_0 + c_1 - 2e^{-\mu_A^U / 2 - \mu_0^U / 2} + 2\sqrt{(1 - e^{-\mu_A^U})(1 - e^{-\mu_0^U})}]}\\
   &\times \sqrt{[c_0 + c_1 - 2e^{-\mu_B^U / 2 - \mu_0^U / 2} + 2\sqrt{(1 - e^{-\mu_B^U})(1 - e^{-\mu_0^U})}]}.
\end{aligned}
\end{equation}

 We also have
\begin{equation}
\bar{N}^{ph}=O^U(\mean{\bar{N}^{ph}}),
\end{equation}  
where $O^U(x)$ is defied in Eq.~\eqref{OU}. The upper bound of the real value of the phase-flip error rate is
\begin{equation}
\bar{e}^{ph}=\bar{N}^{ph}/n_{\mathcal{Z}}.
\end{equation}

Finally, Alice and Bob can calculate the secure final key rate against coherent attacks by Eqs.~\eqref{keyrate_col} and \eqref{keyrate_coh}.

\subsection{Chernoff bound}\label{chernoff}
The Chernoff bound can help us estimate the expected value from their observed values. Let $X_1,X_2,\dots,X_n$ be $n$ independent random samples, detected with the value 1 or 0, and let $X$ denote their sum satisfying $X=\sum_{i=1}^nX_i$. $E$ is the expected value of $X$. We have
\begin{align}
\label{EU}E^U(X)=&\frac{X}{1-\delta_1(X)},
\end{align}
where we can obtain the value of $\delta_1(X)$ by solving the following equations
\begin{align}
\label{delta1}\left(\frac{e^{-\delta_1}}{(1-\delta_1)^{1-\delta_1}}\right)^{\frac{X}{1-\delta_1}}&=\xi,
\end{align}
where $\xi$ is the failure probability.

Besides, we can use the Chernoff bound to help us estimate their real values from their expected values. The observed value, $O$, and its expected value, $Y$, satisfy 
\begin{align}
\label{OU}&O^U(Y)=[1+\delta_1^\prime(Y)]Y,
\end{align}   
where we can obtain the value of $\delta_1^\prime(Y)$ by solving the following equations
\begin{align}
\left(\frac{e^{\delta_1^\prime}}{(1+\delta_1^\prime)^{1+\delta_1^\prime}}\right)^{Y}&=\xi.
\end{align}

\subsection{Signal generation}\label{SIGNAL GENERATION}

\begin{figure}[ht]
\includegraphics[width=\linewidth]{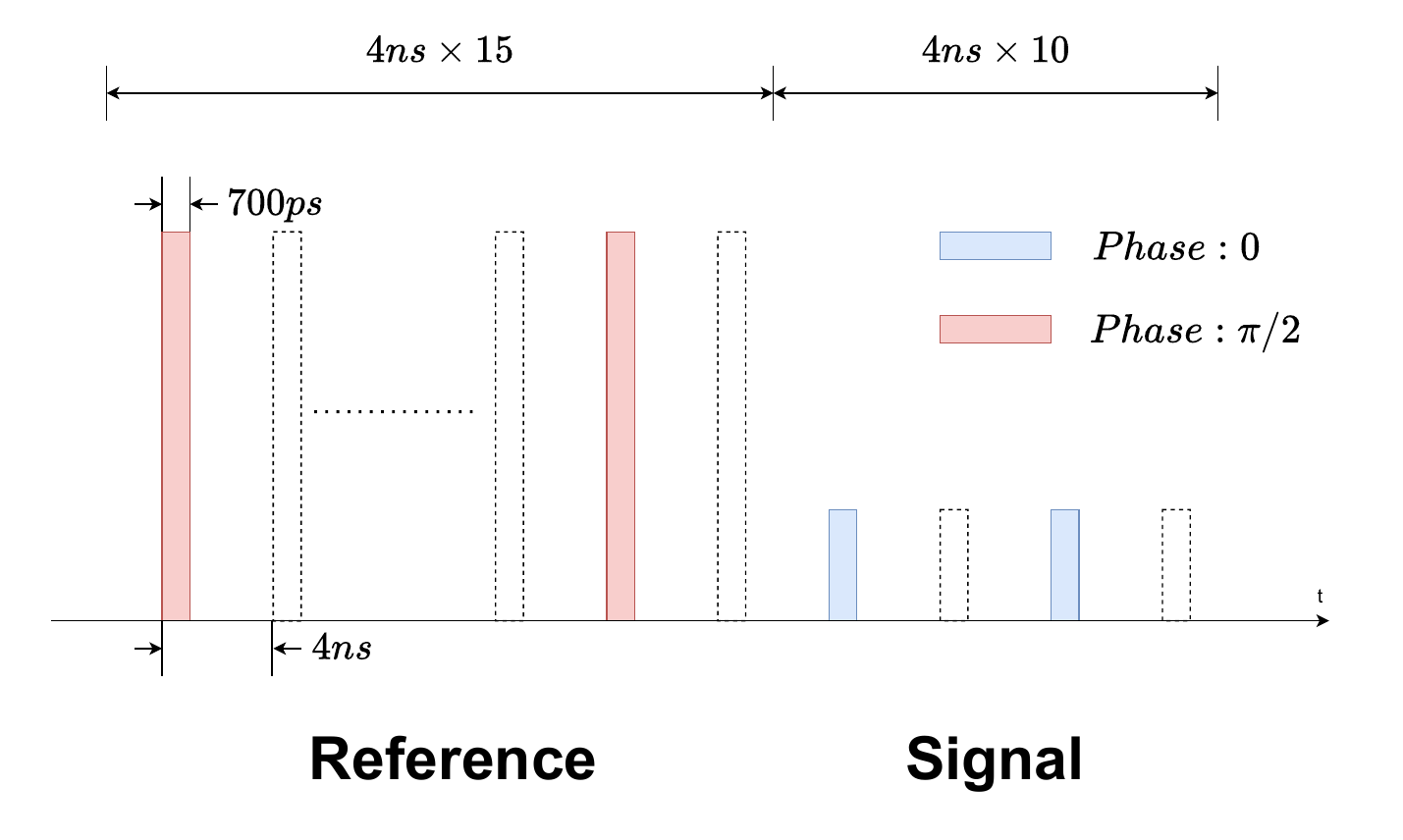}
\caption{\label{fig5}Alice's encoding patten.}
\end{figure}

\begin{figure}[ht]
\includegraphics[width=\linewidth]{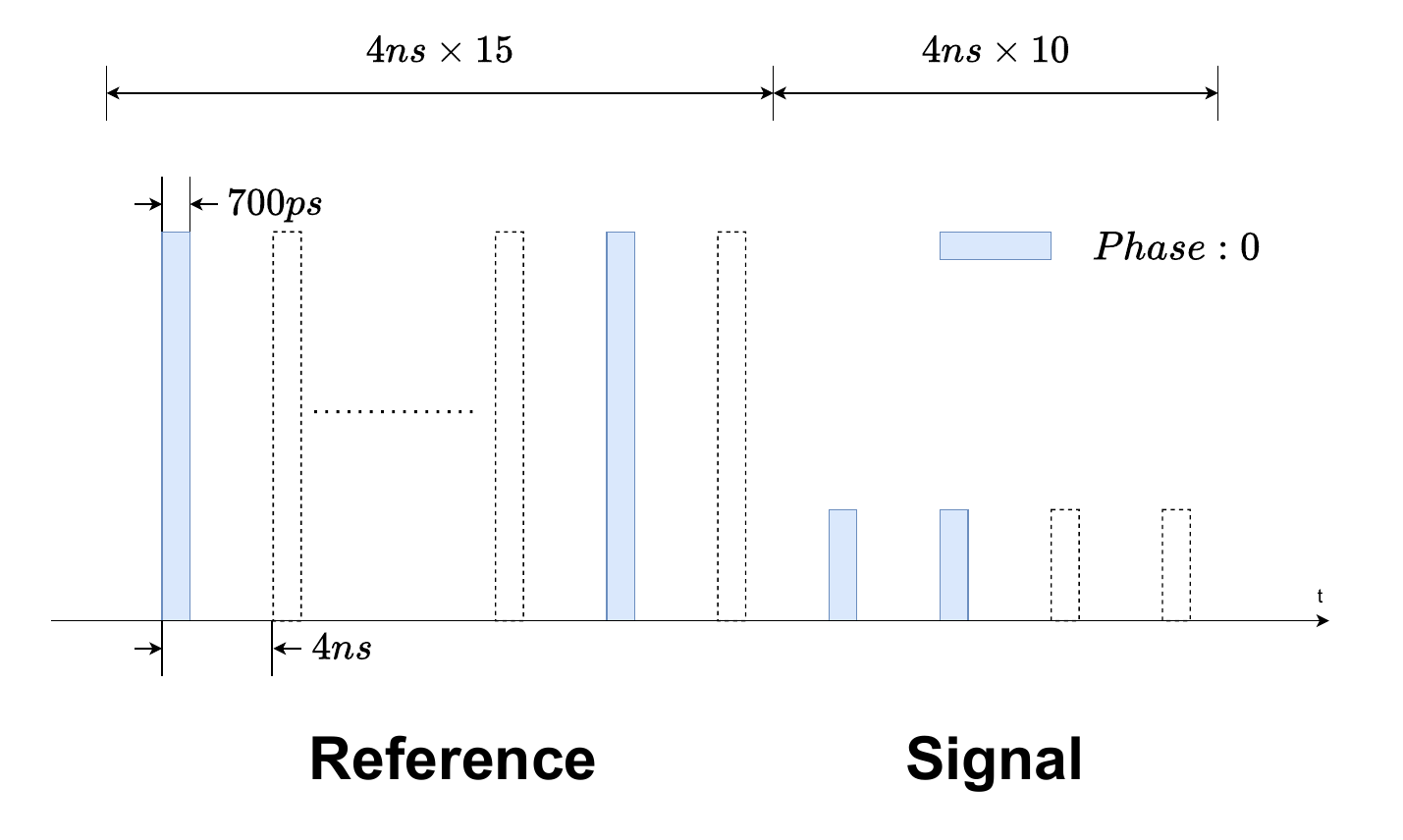}
\caption{\label{fig6}Bob's encoding patten.}
\end{figure}

Alice and Bob use arbitrary waveform generators to generate the modulation signals. Three IMs are used to modulate the continuous wave (CW) lasers to 700 ps pulses width for the ``sending'' bit, or to vacuum state for ``not-sending'' bit. A fourth IM is used to adjust the intensities between the phase reference and the signal pulse. Note that the intensity of the phase reference pulses are modulated with the same time ratio as the signals to ensure the smooth frequency response. As shown in Fig.~\ref{fig5} and Fig.~\ref{fig6}, in each 100 ns time interval, 15 pulses are sent in the first 60 ns, with 700 ps pulse duration in the 4 ns period as the ``reference pulses"; in the next 40 ns, 10 pulses are sent as the ``signal pulses". Random numbers are used to decide ``sending'' or ``not-sending'' for both the quantum signal and phase references. A phase modulator (PM) is inserted at Alice to set the reference pulses phase to $\pi /2$; No additional phase modulation is done at Bob, so the phases of signal pulses and reference pulses at Bob are 0. The signals are further attenuated to the single-photon level by an attenuator.


\begin{figure}[ht]
\centering
\subfigure {\
\begin{minipage}[b]{\linewidth}
        \centering
            \begin{overpic}[scale=0.4]{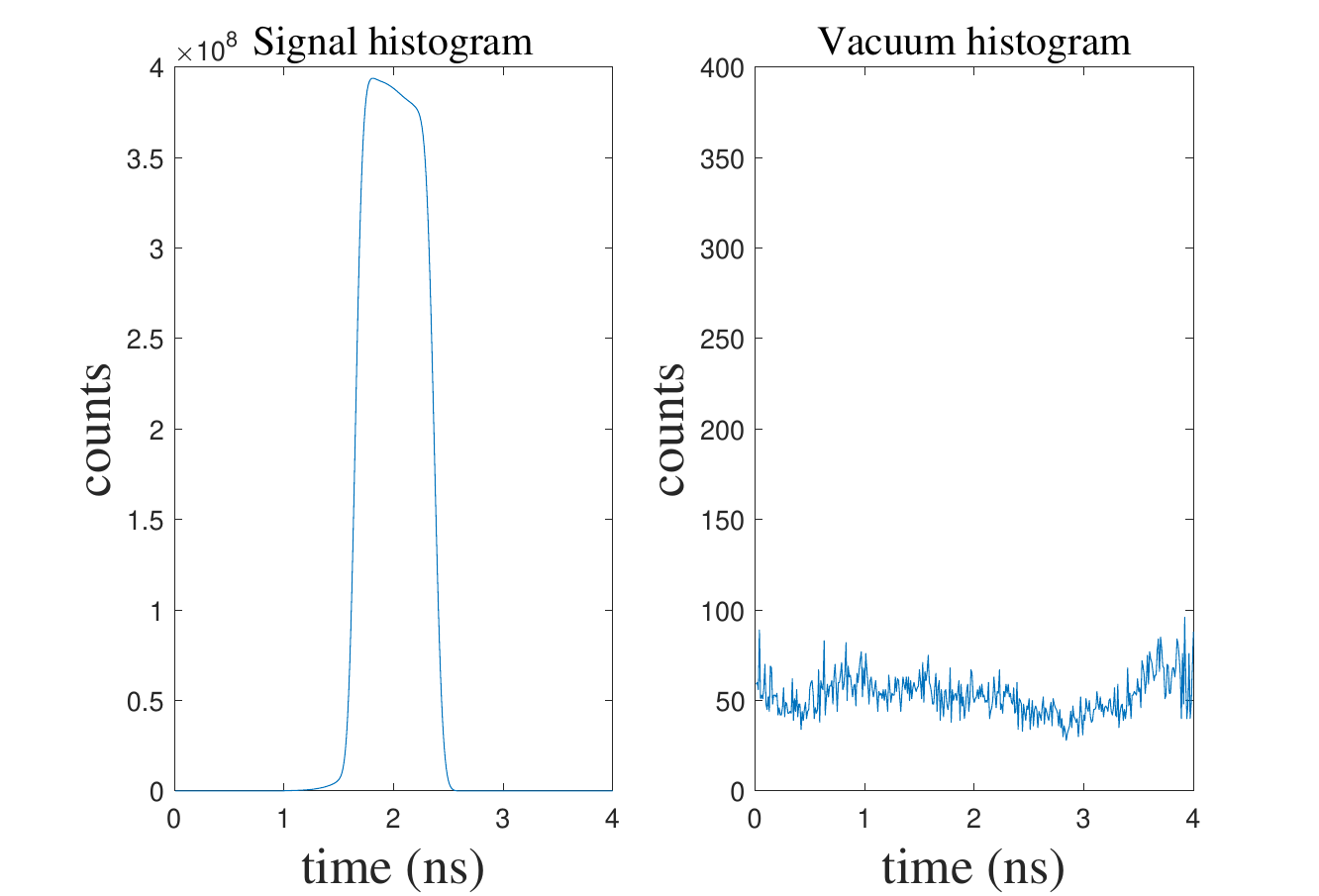}
            \put(1,70){(a)}
        \end{overpic}
    \end{minipage}
    \label{subfig8a}
}
\subfigure {\
\begin{minipage}[b]{\linewidth}
        \centering
           \begin{overpic}[scale=0.6]{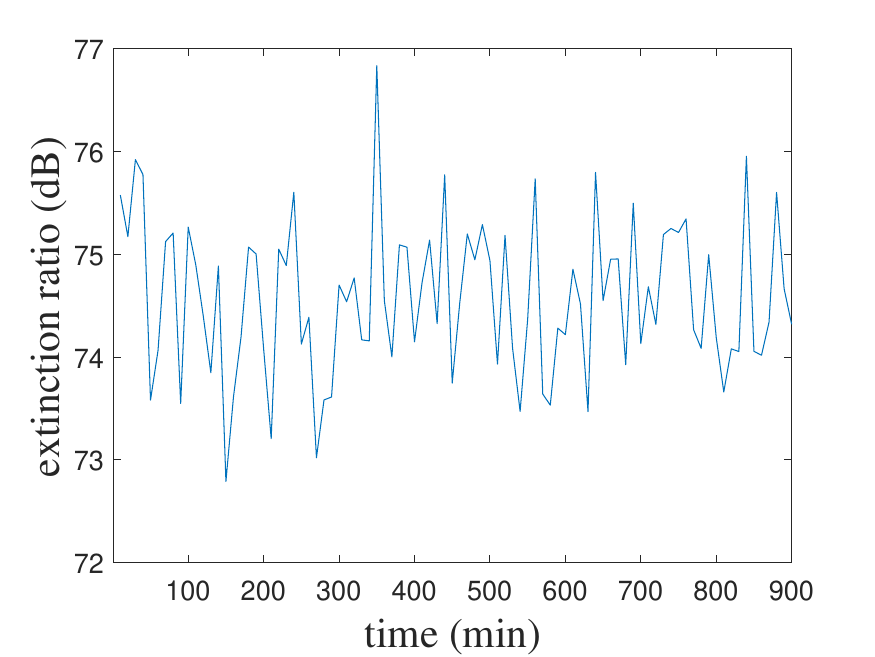}
           \put(1,70){(b)}
        \end{overpic}
    \end{minipage}
    \label{subfig8b}
}
\label{fig8}
\caption{ (a) Schematic diagram of extinction ratio calculation with 900-minute integration time. The left 4 ns temporal window contains the signal histogram, while the right 4 ns window corresponds to the vacuum histogram. The counts in the left 4 ns window with a 700 ps valid window are $2.5717\times10^{10}$, and those in the right 4 ns window are 3639, yielding an extinction ratio of 74.5 dB. (b) Schematic diagram of extinction ratio temporal variation over a 900-minute period, with a 10-minute integration time per each calibration cycle.}
\end{figure}

The extinction ratio between the ``sending'' and ``not-sending'' is critical to the secure key generation in SCS-QKD. To calibrate the high extinction ratio, a low dark-count-rate SNSPD with a dark count rate of 0.14 Hz is used. Random numbers are used as the modulation sequence, with a 1:4 signal-to-vacuum ratio. A TimeTagger is employed to record detection events, with a 10-minute integration time per calibration cycle. We calculate the histograms for signal and vacuum states with a 10 ps bin width. The measured results are sorted comparing with the modulation sequence, yielding the detection counts of signal ($n_{signal}$), and of vacuum ($n_{vacuum}$) with a 700 ps valid time window. The extinction ratio is defined as $ratio = 10\times\log_{10} (n_{signal}\times4/n_{vacuum})$. The measured result of extinction ratio with a 900-minute statistical time is shown in Fig.~\ref{subfig8a}. The left 4 ns temporal window contains the signal histogram, while the right 4 ns window corresponds to the vacuum histogram, the total count of signal in valid time window is $2.5717\times10^{10}$, and that of vacuum is 3639, the extinction ratio is then 74.5 dB. The time dependence of extinction ratio over a 900-minute period is shown in Fig.~\ref{subfig8b}, the extinction ratio is greater than 70 dB during the entire experimental period. 
In calculating the extinction ratio, we did not subtract the SNSPD dark counts, so the actual extinction ratio will be higher.

\subsection{Real-time phase compensation}\label{REAL-TIME PHASE COMPENSATION}

The SCS-QKD protocol requires to compensate for the phase difference of the fibre channel. Specifically, real-time compensation is required to increase the secure key rate in the SCS-QKD protocol.

We developed a real-time phase feedback module to compensate for phase differences in real-time. The module incorporates two signal input channels designed to accept electrical pulse signals from two SNSPDs. The incoming electrical signals are conditioned through a pulse discrimination stage to be converted into standardized TTL-level pulses for FPGA processing. The FPGA-based counting module independently accumulates photon counts from two SNSPDs with an adjustable integration time window. Given the experimental count rates of 1-2 Mcps (mega-counts per second) per detector, the duration of a single counting accumulation cycle ranges from 2 $\upmu$s to 10 $\upmu$s. The registered counts within this temporal window are denoted as $D_{0}$ and $D_{1}$ for the respective detectors. The FPGA data processing module calculates the R-value defined as R = $D_{0}$/($D_{0}$ + $D_{1}$), the sign of R is determined by the relative magnitudes of $D_{0}$ and $D_{1}$. Subsequently, a PID control algorithm generates an output voltage signal based on the deviation between R and its target value. A single PID operation cycle takes approximately 110 ns and the PID controller will not initiate the next computation cycle until the subsequent counting accumulation period is completed. The half-wave voltage of the phase modulator (PM) used in the experiment is 3.05 V, when the PID-computed voltage exceeds 10 V (below 10 V), the actual output voltage equals the PID result minus 6.1 V (plus 6.1 V), ensuring the driving voltage remains within the linear operating regime of the electro-optic crystal. The FPGA's digitized results are precisely output via a 16-bit Digital-to-Analog Converter (DAC).  The DAC features a 50 MSPS bandwidth, and an output range of ±5 V.  We employed a 2$\times$ gain operational amplifier to amplify the electrical signal from the DAC output to drive the PM, and the output value is generated every 1 $\upmu$s, allowing the voltage to smoothly ramp toward the target level, thereby establishing real-time closed-loop phase stabilization through dynamic feedback control.

\subsection{Measurement on pulse correlation}

In practical modulation, the pulse intensity might be correlated with the previous pulses, which might be a result of limited bandwith of the system. This effect is usually called the ``patterning effect''~\cite{Yoshino2018,Lu2021,Roberts:18}. Because the effect is related to the bandwidth response, it is possible in high-speed systems, e.g., QKD systems with clock rate higher than 1 GHz. In this experiment, we set a lower clock rate. Besides, we characterize the pulse correlation with our experimental setup. 

\begin{figure}[ht]
\includegraphics[width=\linewidth]{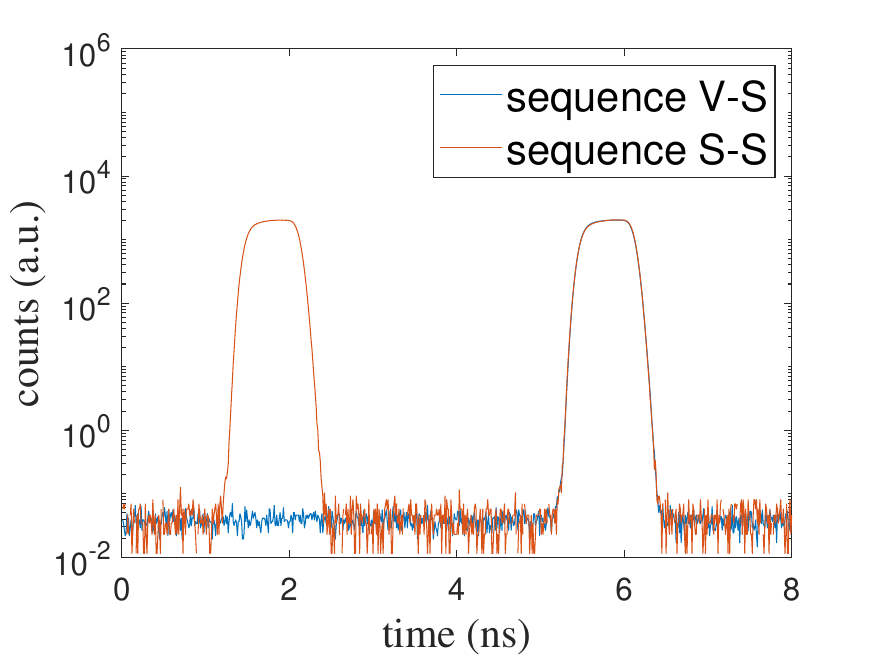}
\caption{\label{fig7}Intensity correlation. The difference of the second pulse between ``VS'' and ``SS'' is calculated to be approximately 0.03\% and the difference of vacuum between ``SV'' and ``VV'' is calculated to be approximately 0.04\%.}
\end{figure}

There are two intensities used in our experiment, denote with ``V'' for the vacuum state, and ``S'' for signal. The number of quantum states transmitted is $9 \times 10^{11}$ times, with the ratio between ``V'' and ``S'' approximately 4 : 1. The exact experimental setup is adopted for test. The SNSPD for the main experiment is used for the measurement. 
The histograms for the closest correlated pulses is calculated using 10 ps bins. In the transmitted pulse sequence, the transmission count for the corresponding group denoted as $T_{i}$, where i represents four corresponding groups of intensity combinations (VS, SS, SV and VV). The detection collection $C_{i}$ is defined as the total detector counts of the second pulse in a two-pulse sequence, with the time window for event counting set to 4 ns. $R_{i}$ = $C_{i}$/$T_{i}$, is defined as the normalized click rate of the second pulse for each sequence. In this test, we process $R_{SS} = 3.5002\times10^{-4}$, $R_{VS} = 3.5011\times10^{-4}$, $R_{SV} = 4.2938\times10^{-8}$, and $R_{VV} = 4.2952\times10^{-8}$. The difference of the second pulse between ``VS'' and ``SS'' is calculated to be approximately 0.03\% and the difference of vacuum between ``SV'' and ``VV'' is calculated to be approximately 0.04\%. The sequence-proportion-normalized distributions of ``VS'' and ``SS'', calculated over a 3600 s interval, are plotted as histograms in Fig.~\ref{fig7}.

\subsection{Measurement on signal intensity fluctuation}\label{SIGNAL INTENSITY FLUCTUATION}

\begin{figure}[ht]
\includegraphics[width=\linewidth]{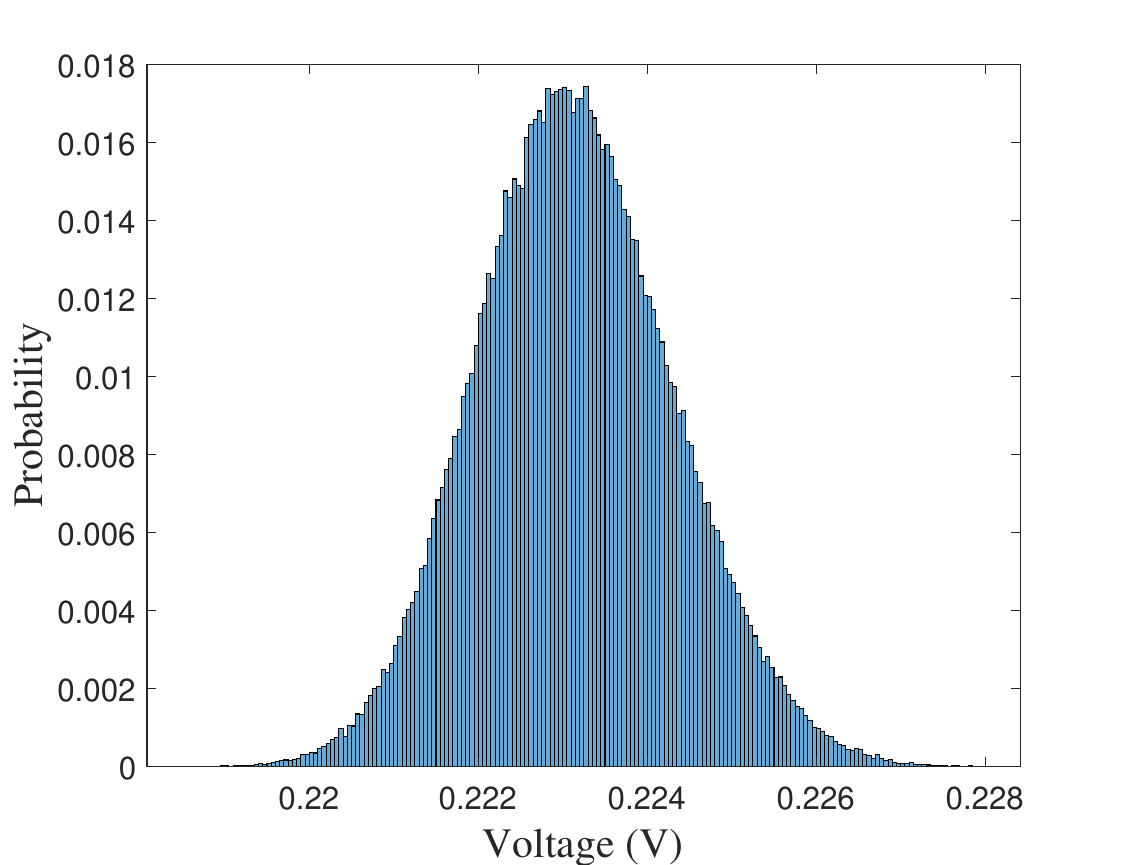}
\caption{\label{fig9}Signal pulse intensity fluctuation. The average pulse intensity of 249,999 pulses is 0.2231 V, with a standard deviation of 0.0012 V.}
\end{figure}

According to the requirements of the SCS protocol, we need to calibrate the upper bound of signal intensity for secure key rate calculation. The modulated signal pulses are converted into electrical signals via a photodiode (PD) and captured using an oscilloscope. The PD has a bandwidth of 40 GHz, the oscilloscope operates at a sampling rate of 20 GSa/s, and the total acquisition duration is 5 ms. The average voltage within 500 ps around the peak of the i-th pulse is taken as the intensity of the i-th pulse ($V_{intensity}^{i}$). Over a 5 ms duration, the intensities of 249,999 pulses exhibit a mean value ($V_{mean}$) of 0.2231 V and a standard deviation ($V_{std}$) of 0.0012 V. Considering 4 times the standard deviation of the signal fluctuation, we conclude that the fluctuation in pulse intensity is $2.15\%$. The statistical histogram of the $V_{intensity}^{i}$ is shown in the Fig.~\ref{fig9}.

\subsection{Detailed experimental parameters and results}\label{DETAILED EXPERIMENTAL PARAMETERS AND RESULTS}

The parameters used in our experiment are summarized in Tab.~\ref{tab:table1}.  For 0 km, 50.5 km and 101.1 km, the fibre length from Alice to Charlie and from Bob to Charlie are set to the same; for 25.3 km and 75.7 km, the fibre length from Alice to Charlie is 25.3 km longer than that from Bob to Charlie. The ``Total distance'' is the total fibre length between Alice and Bob, and the $Distance_{A(B)-C}$ is the fibre length between Alice (Bob) and Charlie. The ``Total Loss'' is the total fibre loss between Alice and Bob, and the $Loss_{A(B)-C}$ is the fibre loss between Alice (Bob) and Charlie. $\mu_{A (B)}$ is the mean intensity of the coherent state of signal pulse sent by A (B). $\mu_{A (B)}^{U}$ is the upper bound of the $\mu_{A (B)}$. The ``sending'' probability is denoted by $P_{S}$. Finally, we listed the experimental time, the QKD system frequency, and the equivalent system frequency (considering phase reference pulses) in the experiment.

\begin{table}[ht]
\caption{Experimental parameters.}
\label{tab:table1}
\begin{ruledtabular}
\begin{tabular}{lccccc}
Total distance (km)&0&25.3&50.5&75.7&101.1\\
\hline
$Distance_{A-C}$ (km)&0&25.3&25.3&50.5&50.5\\
$Distance_{B-C}$ (km)&0&0&25.2&25.2&50.6\\
\hline
Total Loss (dB)&0&4.25&8.48&12.71&16.98\\
\hline
$Loss_{A-C}$ (dB)&0&4.25&4.25&8.48&8.48\\
$Loss_{B-C}$ (dB)&0&0&4.23&4.23&8.5\\ 
\hline
$\mu_{A} $&0.0174&0.0137&0.0061&0.0045&0.0019\\ 
$\mu_{B} $&0.0348&0.0137&0.0122&0.0045&0.0038\\ 
$\mu_{A}^{U} $&0.0178&0.0140&0.0062&0.0046&0.0019\\ 
$\mu_{B}^{U} $&0.0356&0.0140&0.0124&0.0046&0.0039\\ 
\hline
Experimental Time (h)&\multicolumn{4}{c}{1.01}&3.13\\
Extinction ratio S/V (dB)&\multicolumn{5}{c}{\textgreater 70}\\
``Sending'' prob. $P_{S}$&\multicolumn{5}{c}{0.2}\\
System Freq. (MHz)&\multicolumn{5}{c}{250}\\
Effective Freq. (MHz)&\multicolumn{5}{c}{100}\\
\end{tabular}
\end{ruledtabular}
\end{table}

The characteristics of superconducting nanowire single photon detectors (SNSPDs) are presented in Tab.~\ref{tab:table2}. Tab.~\ref{tab:table3} summarizes the efficiency of a series of components with respect to each user’s input port in Charlie’s module. The optical elements include the polarization beam splitters (PBSs), the 50:50 Polarization maintaining beam splitter (PMBS) and a phase modulator which is only set in Bob's side. 

\begin{table}[ht]
\caption{\label{tab:table2}Characteristics of Charlie’s detectors $D_{0}$ and $D_{1}$.}
\begin{ruledtabular}
\begin{tabular}{lcc}
\textrm{Detector}&\textrm{Efficiency}&\textrm{Dark count}\\
\hline
$D_{0}$&68.2\%&8.4 Hz\\ 
$D_{1}$&70.5\%&7.5 Hz\\
\end{tabular}
\end{ruledtabular}
\end{table}

\begin{table}[ht]
\caption{\label{tab:table3}Efficiencies of Charlie’s components.}
\begin{ruledtabular}
\begin{tabular}{lcc}
&\textrm{Alice}&\textrm{Bob}\\
\hline
\textrm{Polarization beam splitter}&92.7\%&92.1\%\\ 
\textrm{Phase modulator}&\textrm{n/a}&49.6\%\\
\textrm{50:50 Polarization maintaining beam splitter}&45.7\%&45.9\%\\
\end{tabular}
\end{ruledtabular}
\end{table}

Table~\ref{tab:table4} summarizes the raw data and experimental results used for secure key rate (SKR) calculation. $N$ is the total number of transmitted quantum pulses. The numbers of pulses that Alice and Bob send are labelled as “Sent-AB”, where“A” (“B”) is “1” or “0”, indicating the intensity Alice (Bob) for “sending” or “not-sending”. The numbers of detections are listed as “Detected-AB-ch”, where "ch" indicates the detection channel. The higher vacuum detection rate (``Detected-00-ch'', approximately 50× higher) is due to the higher dark count noise of SNSPD during the measurement. QBER indicates the error rate that Alice and Bob both decide to send.

\begin{table*}[ht]
\caption{\label{tab:table4}Experimental results at various quantum link fibre lengths}
\begin{ruledtabular}
\begin{tabular}{lccccc}
Total length (km)&0&25.3&50.5&75.7&101.1\\
\hline
$N_{total}$&$3.63\times 10^{11}$&$3.63\times 10^{11}$&$3.63\times 10^{11}$&$3.63\times 10^{11}$&$1.126\times 10^{12}$\\
$R$ (per pulse)&$6.47\times 10^{-4}$&$1.38\times 10^{-4}$&$6.60\times
10^{-5}$&$1.28\times 10^{-5}$&$3.58\times 10^{-7}$\\
$R$ (bps)&$6.47\times 10^{4}$&$1.38\times 10^{4}$&$6.60\times 10^{3}$&$1.28\times 10^{3}$&$3.58\times 10^{1}$\\
\textrm{Sent-00}&231594000000&231594000000&231594000000&231594000000&718388000000\\
\textrm{Sent-01}&58806000000&58806000000&58806000000&58806000000&182412000000\\
\textrm{Sent-10}&58806000000&58806000000&58806000000&58806000000&182412000000\\
\textrm{Sent-11}&13794000000&13794000000&13794000000&13794000000&42788000000\\
\textrm{Detected-00-ch0}&1638729&34857&35275&48849&29694\\
\textrm{Detected-00-ch1}&4149624&42620&36922&54023&27847\\
\textrm{Detected-01-ch0}&301133352&93778052&40806039&11213924&11733761\\
\textrm{Detected-01-ch1}&312923338&96053698&41754462&11452902&12213595\\
\textrm{Detected-10-ch0}&330999032&89590603&39004871&11920527&11622299\\
\textrm{Detected-10-ch1}&346403215&92394051&40119263&12256893&12178754\\
\textrm{Detected-11-ch0}&6117941&2737041&1002113&305360&557067\\
\textrm{Detected-11-ch1}&306850280&88868203&39536414&11998373&12903180\\
\textrm{QBER}&1.95\%&2.99\%&2.47\%&2.48\%&4.14\%\\
\end{tabular}
\end{ruledtabular}
\end{table*}

\end{document}